\documentclass[a4paper]{jpconf}
\usepackage{graphicx}
\begin{document}
\title{Gravitational Wave Background from Neutron Star Phase Transition
for a new class of equation of state}
\author{Jos\'e Carlos N. de Araujo$^1$ and Guilherme F. Marranghello$^2$}
\address{$^1$Instituto Nacional de Pesquisas Espaciais, S\~ao Jos\'e dos Campos/SP, Brazil}
\address{$^2$Universidade Federal do Pampa, Bag\'e/RS, Brazil}
\ead{jcarlos@das.inpe.br}

\begin{abstract}
We study the generation of a stochastic gravitational wave (GW)
background produced by a population of neutron stars (NSs) which go
over a hadron-quark phase transition in its inner shells. We obtain,
for example, that the NS phase transition, in cold dark matter
scenarios, could generate a stochastic GW background with a maximum
amplitude of $h_{\rm BG} \sim 10^{-24}$, in the frequency band
$\simeq 20-2000\, {\rm Hz}$ for stars forming at
redshifts of up to $z\simeq 20.$ We study the possibility of
detection of this isotropic GW background by correlating signals of
a pair of `advanced' LIGO observatories.
\end{abstract}

\section{Introduction}

It is widely accepted that neutron stars (NSs) are born with fast
rotating angular velocities. Due to magnetic torques, however, the
NS periods could well be spun down. This spin-down causes a
reduction in the centrifuge force and, consequently, the central
energy density of the NSs increases. Those stars, born with
densities close to that of the quark deconfinement, may undergo a
phase transition forming a strange quark matter core. It is worth
stressing, however, that it is not a common sense that such a
transition really takes place.

As a consequence of such a putative phase transition, the
stars could suffer a collapse which could excite
mechanical oscillations. The great amount of energy generated in
this process, $\Delta E\sim 10^{53}\,ergs$ ($\sim 0.1\,
M_{\odot}c^{2}$) \cite{marr1}, could be dissipated, at least
partially, in the form of gravitational waves (GWs), which are
studied in the present work.

We adopted in the present paper the history of star formation derived by Springel \& Hernquist \cite{springel}, who employed hydrodynamic simulations of structure formation in a $\Lambda$ cold dark matter ($\Lambda$CDM) cosmology. These authors study the history of cosmic star formation from the ``dark ages", at redshift  ${\rm z} \sim 20$, to the present.

Besides the reliable history of star formation by Springel \& Hernquist, we consider the
role of the parameters $\alpha_{i=1,2}$, which gives the fraction of the progenitor mass which forms the remnant NSs. We consider that the remnant mass is given as a function of the progenitor mass, namely $M_r=\alpha_1m+\alpha_2$. Later on we justify why $M_r$ is written in this way. Recall that a given initial mass function (IMF) refers to the distribution function of the stellar progenitor mass, and also to the masses of the remnant compact objects left as a result of the stellar evolution.

In the present study we have adopted a stellar generation with a
Salpeter IMF, which is consistent with Springel \& Hernquist,
since they show that population II stars could have been formed at
high redshift too. We then discuss briefly what conclusions would be drawn
whether (or not) the stochastic background studied here is detected
by GW observatories such as LIGO and VIRGO.

The paper is organized as follows. Section 2 deals with the NS EOSs
adopted in our study; in Section 3 we consider the IMF and the NS
masses; Section 4 deals with the GW production; in Section 5 we
present the numerical results and discussions; Section 6 the
detectability of the background of GWs are considered and finally in
Section 7 we present the conclusions.

\section{Neutron stars and the nuclear matter equation of state}

We consider in the present study a particular model developed
by Taurines et al \cite{taurines} for a new class of parameterized
field-theoretical model described by a Lagrangian density, where the
whole baryon octet is coupled to scalar and vector fields through
parameterized coupling constants. The free lepton fields contributes
to the electrical equilibrium in the NS matter.

Using a parametrization for the baryon-meson coupling constants,
this model describes a wide range of NS parameters, such as a
maximum mass ranging from (very low values) $M=0.66M_\odot$ up to
$M=2.77M_\odot$, and the corresponding radii that vary in the range
of $8<R<13\, km$. Each pair in the mass-radius relation is
associated with a different parametrization of the EOS (see
\cite{taurines} for further details). Of course, each value of such
parametrization represents different values of nuclear matter
properties.

Another scenario to be discussed concerns the strange quark star.
The model used in this analysis is the MIT bag model \cite{mit}.
The effects of pressure difference in the interior and exterior
regions of the bag are summarized in the bag constant.

Once the transition to quark-gluon matter occurs, the weak
interaction processes for the quarks u, d and s
\begin{equation}
u + s \rightarrow d + u
\end{equation}
and
\begin{equation}
d + u \rightarrow u + s
\end{equation}
will take place, and a rapid transition occurs with a consequent
gravitational micro-collapse. As a result, a huge amount of energy
is dissipated, some in the form of GWs produced by quasi-normal
modes excitation.

\section{Initial mass function and the neutron star masses}

The calculation of the GW background from NS phase transition requires the knowledge
of the distribution function of stellar masses, the so called stellar initial mass function (IMF), $\phi(m)$. Here the Salpeter IMF is adopted, namely

\begin{equation}
\phi(m) = A m^{-(1+x)},
\end{equation}

\noindent where $A$ is the normalization constant and $x=1.30$ (our fiducial value). The
normalization of the IMF is obtained through the relation

\begin{equation}
\int_{m_{\rm l}}^{m_{\rm u}} m\phi(m)dm = 1,
\end{equation}

\noindent where we consider $m_{\rm l} = 0.1\, {\rm M}_{\odot}$ and $m_{\rm u}=125\, {\rm
M}_{\odot}$. For further details we refer the reader to, e.g., \cite{araujo2002}.

It is worth mentioning that concerning the star formation at high redshift, the IMF could
be biased toward high-mass stars, when compared to the solar neighborhood IMF, as a result
of the absence of metals \cite{bromm99,bromm02}.

On the other hand, in the study by Springel and Hernquist, besides a high-mass star
formation, it is shown that the population II stars, whose IMF could well be of the
Salpeter's type, could start forming around redshift 20 or higher.

In the present study we consider this population II studied by these authors.
Then, for the standard IMF, the mass fraction of NSs produced as remnants of the stellar evolution is

\begin{equation}
f_{\rm NS} = \int_{\rm m_{min}}^{m_{\rm u}}M_{\rm r}\phi(m)dm,
\end{equation}

\noindent where $m_{\rm min}$ is the minimum stellar mass capable of producing a NS at the
end of its life, and $M_{\rm r}$ is the mass of the remnant NS. Stellar evolution
calculations show that the minimal progenitor mass to form NSs is $m_{\rm min}=8 {\rm
M}_{\odot}$, while the maximum progenitor mass is $m_{\rm max}= 25 - 40 {\rm M}_{\odot}$
(see, e.g., \cite{timmes}).

We do not intend to discuss in the present paper stellar evolution scenarios related to the
formation of NSs. Many works can be found in the literature concerning the study of stellar
evolution, supernova and NS formation \cite{janka}. However, we need to formulate a relation
between the progenitor star mass $m$ and the remnant NS mass $M_{\rm  r}$. For the remnant,
$M_{\rm  r}$, we take

$$M_{\rm r}=\alpha_1\, m+\alpha_2 \, ,$$

\noindent where $\alpha_1$ and $\alpha_2$ are constants; the first one is dimensionless and the second one is given in solar masses. As the results for stellar evolution are not yet fully determined, we studied this parametrization in three different scenarios for the values of $\alpha_1$ and $\alpha_2$ which would represent the distribution of NS mass as function of the progenitor mass (see Table 1 and the following sections). For example, considering NSs formed from progenitors with $m=20M_\odot$ we would have remnants with masses $M_r=$ 1.375$M_\odot$, 1.5$M_\odot$ and 2.7$M_\odot$. These parameters describe sharper or softer distribution of NS masses around the standard value of mass $M=1.4M_\odot$ . Note that we are considering in the end that $M_{\rm r}$ is independent of the redshift.

\begin{table}
\caption {The values of $\alpha_{i=1,2}$ which determines the fraction of the
  NS remant.}
\begin{center}
\begin{tabular}{lccc}
\hline
 $\alpha_1$ & 1/32 & 1/50 & 1/17 \\
 $\alpha_2 \,(M_\odot) $ & 3/4  & 1.1  & 0.53 \\
\hline
\end{tabular}
\end{center}
\end{table}

With these considerations at hand, the mass fraction of NSs reads up to $f_{\rm
NS}=10^{-2}$ for $x=1.30$, while the fraction of NSs that undergoes a phase transition
($f_{\rm NS}^{pt}$) can drop down to values $\ll 1$ (see, e.g., \cite{marr2}).

\section{Gravitational wave production}

The dimensionless amplitude of the Gravitational Wave Background
from Neutron Star Phase Transition can be given by

\begin{equation}
h_{\rm BG}^{2} = {1 \over \nu_{\rm obs}}\int h_{\rm NS}^{2} dR,
\end{equation}

\par\noindent (see \cite{araujo2000,araujo2005}).

The micro-collapse produces GWs at frequency $\nu$ of the NS f-mode,
and dimensionless amplitude given by \cite{ander1998}

\begin{equation} \label{h}
h_{NS} \simeq 1\times10^{-19}\left(\frac{E}{M_\odot
c^2}\right)^{1/2} \left(\frac{2 kHz}{\nu}\right)^{1/2}\left(\frac{1
Mpc}{d_{\rm L}}\right)
\end{equation}

\noindent where $E$ is the available pulsation energy, and $d_{\rm
L}$ is the luminosity distance to the source.

Recall that the energy available to excite the pulsating modes is
directly related to the EOS adopted.

For the differential rate of NS micro-collapse we have
\begin{equation}
dR_{\rm NS} = \dot\rho_{\star}(z) {dV\over dz} f_{\rm NS}^{pt}
\phi(m)dmdz,
\end{equation}
where $\dot\rho_{\star}(z)$ is the star formation rate (SFR) density
(in ${\rm M}_{\odot}\,{\rm yr}^{-1}\,{\rm Mpc}^{-3}$),  $dV$ is the
comoving volume element, and $f_{\rm NS}^{pt}$, as already
mentioned, is the fraction of NSs which may undergo a phase transition
forming a strange quark matter core.

For the SFR density, we adopt the one derived by Springel and
Hernquist (see, \cite{springel} for details), namely

\begin{equation}
\label{eqnfit} \dot\rho_\star(z)= \dot\rho_m\,
\frac{\beta\exp\left[\Delta(z-z_m)\right]}
{\beta-\Delta+\Delta\exp\left[\beta(z-z_m)\right]},
\end{equation}

\noindent where $\Delta= 3/5$, $\beta=14/15$, $z_m=5.4$ marks a
break redshift, and $\dot\rho_m= 0.15\,{\rm M}_\odot{\rm
yr}^{-1}{\rm Mpc}^{-3}$ fixes the overall normalization.

It is worth mentioning that these authors employed hydrodynamic
simulations of structure formation in a $\Lambda$CDM cosmology with
the following parameters: $\Omega_{\rm M}=0.3$,
$\Omega_{\Lambda}=0.7$, Hubble constant $H_{0}=100\; h\; {\rm
km\;s^{-1}\;Mpc^{-1}}$ with $h=0.7$, $\Omega_{\rm B}=0.04$, and a
scale-invariant primordial power spectrum with index $n=1$,
normalized to the abundance of rich galaxy clusters at present day
($\sigma_{8}=0.9$).

Another relevant physical quantity associated with the GW
background, produced by the first stars, is the closure energy
density per logarithmic frequency span, which is given by

\begin{equation}
\Omega_{\rm GW} = {1\over \rho_{\rm c}} {d\rho_{\rm GW}\over d\log
\nu_{\rm{obs}}}.
\end{equation}

In the next section we present the numerical results and
discussions, which come mainly from the equation for ${h_{\rm BG}}$.

\section{NUMERICAL RESULTS AND DISCUSSIONS}

In order to cover a wide number of parameters we present, in Table 2, the models considered in our study. In the first column appears the name of the model; in the second and third columns we present different combinations of the parameters $\alpha_{1}$ and $\alpha_{2}$, which imply in different ways to calculate the NS remnant mass for a given IMF; in the fourth column the mass range of the progenitor star; in the fifth column the NS remnant mass; in the sixth column the NS redshift formation, and finally in the seventh column the observed frequency, which are obtained via the empirical formulae given in reference \cite{benhar}.

These parameters are directly related to the number of NSs that goes over the phase transition. According to these NS models, only a fraction of them develops a core composed by deconfined quarks and gluons. When we set the mass range of the progenitor mass, we are also setting the mass of the remnant object. Stars with higher mass reach central densities that are high enough to develop the quark core, while smaller stars do not. The available energy for GW emission generated in the collapse process is also implicit in the mass of the remnant star. The more massive is the star, greater is its core and greater is the energy difference between the neutron and the hybrid star. We have found that stars with baryonic mass $M_b=2M_\odot$ develops a core as large as $R=7km$ and can generate as much as $3\times10^{53}erg$ of energy, while stars with baryonic mass $M_b=1M_\odot$ develops a small core with $R=1km$ and has ten times less energy available from the transition. As we do not know for sure the amount of this energy that is driven in each mode and in order to simplify our calculations we have adopted the standard medium value of 0.01$M_\odot$ for the released energy.

\begin{figure}[h]\hspace{3cm}
\includegraphics[width=20pc]{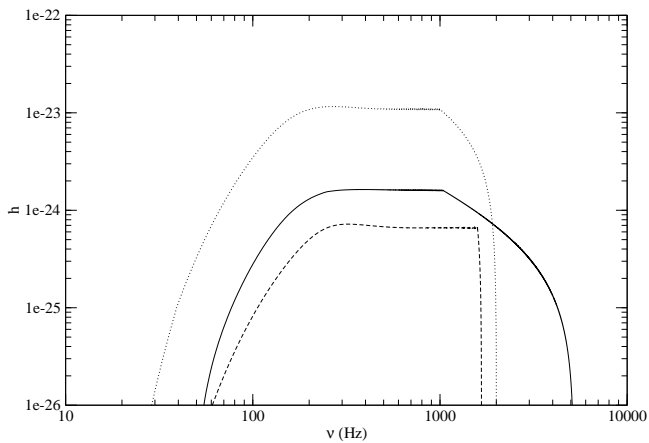}
\caption{\label{2}
The dimensionless
amplitude for the GW background of  stellar black hole formation
(solid line), the whole ensemble of NSs
undergoing phase transition (dotted line) and 1 \% of the NSs doing
so (dashed line), for $0<z<20$.}
\end{figure}

In figure \ref{2} we compare the background of GWs generated by NSs,
which undergo phase transition, with that by the BH formation, which
undergo quasi-normal mode instability (see \cite{araujo2004} for
details). In particular, for the NSs we consider two situations. We
refer the reader to Table 2 to see the parameters adopted in these
calculations.

\vskip 12pt

\begin{table*}
\caption {Models description: mass function constants,
$\alpha_{1,2}$, progenitor, $m$, and remnant, $M$, mass ranges,
redshift, $z$, and the observed frequency, $\nu$.}
\vskip 12pt
\begin{center}
\begin{tabular}{ccccccc}
\hline Model & $\alpha_1$ & $\alpha_2$ & $\Delta m (M_\odot)$ &
$\Delta M (M_\odot)$ &
$\Delta z$ & $\Delta\nu (Hz)$ \\
\hline
A  & $1/32$ & $3/4 $ & 8.0-40.00  & 1.00-2.00 & 0-20 & 20-2000 \\
B  & $1/32$ & $3/4 $ & 14.4-16.32 & 1.20-1.26 & 0-20 & 31-1666 \\
C  & $1/32$ & $3/4 $ & 14.4-16.32 & 1.20-1.26 & 0-20 & 75-1666 \\
D  & $1/32$ & $3/4 $ & 20.8-22.72 & 1.40-1.46 & 0-20 & 27-1428 \\
E  & $1/50$ & $1.1 $ & 8.0-40.00  & 1.26-1.90 & 0-20 & 21-1587 \\
F  & $1/50$ & $1.1 $ & 15.0-18.00 & 1.40-1.46 & 0-20 & 27-1428 \\
G  & $1/17$ & $0.53$ & 8.0-25.00  & 1.00-2.00 & 0-20 & 20-2000 \\
\hline
\end{tabular}
\end{center}
\end{table*}

We consider in figure \ref{2} two versions of the model B: one with its
original values, namely, $f_{\rm NS}^{pt} = 0.01$ and the other one,
just as an example, with  $f_{\rm NS}^{pt} = 1$. Obviously, even the
most favorable values for nuclear and subnuclear matter coupling
constants could not make $f_{\rm NS}^{pt}$ even tend to 1, but this
model is included just as a matter of comparison, as it would
represent an upper limit. In both cases we consider E=0.01, i.e.,
only a tenth of the energy generated in phase transition goes to the
pulsating mode.

Note that the background of GWs generated by the NSs would have an
amplitude greater than that generated by the BHs only if a
considerable fraction of the NSs formed undergo phase transition.

We compare how different laws to calculate the NSs
mass, the remnant mass, for a Salpeter IMF, modify the spectrum of
the background of GWs generated. We obtain that there is no significantly
difference in the background for the three cases considered.

It is worth mentioning that a detailed discussion of the models 
presented in Table 2, among other issues, will be considered in another publication to 
appear elsewhere.

\section{DETECTABILITY OF THE BACKGROUND OF GRAVITATIONAL WAVES}

The background predicted in the present study cannot be detected by
single interferometric detectors, such as VIRGO and LIGO
(even by advanced ones). However, it is possible to correlate the
signal of two or more detectors to detect the background that we
propose exists.

To assess the detectability of a GW signal, one must evaluate the
signal-to-noise ratio (S/N), which for a pair of interferometers is
given by (see, e.g., \cite{flanagan,allen1999})

\begin{equation}
\label{sn} {\rm (S/N)}^2=\left[\left(\frac{9 H_0^4}{50\pi^4} \right)
T \int_0^\infty d\nu \frac{\gamma^2(\nu)\Omega^2_{GW}(\nu) } {\nu^6
S_h^{(1)}(\nu) S_h^{(2)}(\nu)} \right]
\end{equation}

\noindent where $ S_h^{(i)}$ is the spectral noise density, $T$ is
the integration time and $\gamma(\nu)$ is the overlap reduction
function, which depends on the relative positions and orientations
of the two interferometers. For the $\gamma(\nu)$ function we refer
the reader to \cite{flanagan}, who was the first to calculate a
closed form for the LIGO observatories.

Here we consider, in particular, the LIGO interferometers. Their
spectral noise densities have been taken from \cite{owen} - who in
turn obtained them from Thorne, by means of private communication.
It is worth mentioning that although a ten years old reference has been
cited, it presents analytical equations for the spectral noise densities, which are
in complete agreement with many recent papers involving detectability of GWs by the LIGOs.

In Table 2, as already mentioned, we present the models considered in our 
study. We show in Table 3 the S/N for the models of Table 2 for the three
different LIGO generations and for different values of $f^{pt}_{NS}$.

\begin{table*}
\caption {For the models of Table 2 with different values of
$f^{pt}_{NS}$, we present the S/N for pairs of initial, enhanced and advanced LIGO
observatories for one year of observation. We consider in the calculations $E = 0.01
M_{\odot}c^{2}$.}
\vskip 12pt
\begin{center}
\begin{tabular}{ccccc}
\hline
& & &  S/N &   \\
Model & $f^{pt}_{NS}$ & Initial LIGO  & Enhanced LIGO  & Advanced LIGO \\
\hline
A & 1.0  & $3.6\times 10^{-3}$  & $1.8\times 10^{-1}$ & $4.3\times 10^{-1}$  \\
B & 0.01 & $6.5\times 10^{-6}$  & $2.6\times 10^{-4}$ & $5.0\times 10^{-4}$  \\
C & 0.01 & $6.5\times 10^{-6}$  & $2.6\times 10^{-4}$ & $5.0\times 10^{-4}$  \\
D & 0.01 & $7.5\times 10^{-6}$  & $3.1\times 10^{-4}$ & $6.5\times 10^{-4}$  \\
E & 1.0  & $2.8\times 10^{-3}$  & $1.4\times 10^{-1}$ & $3.3\times 10^{-1}$  \\
F & 1.0  & $1.7\times 10^{-5}$  & $7.0\times 10^{-4}$ & $1.5\times 10^{-3}$  \\
G & 1.0  & $2.7\times 10^{-3}$  & $1.4\times 10^{-1}$ & $3.3\times 10^{-1}$  \\
\hline
\end{tabular}
\end{center}
\end{table*}

As shown in Table 3, the signal-to-noise ratio for all models
studied is lower than one, even for an advanced LIGO. Therefore,
contrary to the claim of \cite{sigl} such a putative GW background
would hardly be detected.

Note that the signal-to-noise ratio, for given IMF, $\alpha_{i}$,
and integration time, depends on $f_{\rm NS}^{pt}$ and $E$ as
follows
\begin{equation}
{\rm (S/N)} \propto f_{\rm NS}^{pt} E \, ;
\end{equation}

\noindent and it also depends on the SFR density in a more
complicated way, namely, through an integral involving the redshift
$z$. The higher the star formation rate, the higher the signal-to
noise ratio will be.

Just a matter of comparision, even considering models
with $f_{\rm NS}^{pt}=1$, a signal-to-noise ratio significantly
greater than one, for the advanced LIGO, would be possible either
the SFR density would be much greater than that by Springel \&
Hernquist, or the energy generated in the phase transition were
almost completely channeled to excite the f- mode. Obviously, an
optimist combination of the SFR density and the energy channeled to
the f-mode would also render the same.

\section{CONCLUSIONS}

We present here a study concerning the generation of GWs produced
from a cosmological population of NSs. These stars may undergo a
phase transition if born close to the transition density, suffering
a micro-collapse and exciting quasi-normal modes.

We show that a detectable background is possible only if the SFR
density is much greater than that predicted by Springel \& Hernquist
or if the energy generated in the phase transition is almost completely
channeled to excite the f- mode. Obviously, a too optimistic
combination of these two possibilities could do the same.

\ack{JCNA would like to thank CNPq and Fapesp for financial support.}

\end{document}